# Negative Zero-Point-Energy Parameter in the Meyer-Miller Mapping Model for Nonadiabatic Dynamics

*Xin He[†], Zhihao Gong[†], Baihua Wu, and Jian Liu\**

Beijing National Laboratory for Molecular Sciences, Institute of Theoretical and Computational Chemistry, College of Chemistry and Molecular Engineering,

Peking University, Beijing 100871, China

AUTHOR INFORMATION

**Corresponding Author**

\* Email: jianliupku@pku.edu.cn

**Author Contributions**

† X. H. and Z. G. contributed equally.

**ABSTRACT**. The celebrated Meyer-Miller mapping model has been a useful approach for generating practical trajectory-based nonadiabatic dynamics methods. It is generally assumed that the zero-point-energy (ZPE) parameter is positive. The constraint implied in the conventional Meyer-Miller mapping Hamiltonian for an $F$-electronic-state system actually requires $\gamma \in (-1/F, \infty)$ for the ZPE parameter for each electronic degree of freedom. Both negative and positive values are possible for such a parameter. We first establish a rigorous formulation to construct exact mapping models in the Cartesian phase space when the constraint is applied. When nuclear dynamics is approximated by the linearized semiclassical initial value representation, a negative ZPE parameter could lead to reasonably good performance in describing dynamic behaviors in typical spin-boson models for condensed-phase two-state systems, even at challenging zero temperature.

**Abstract Graphic**

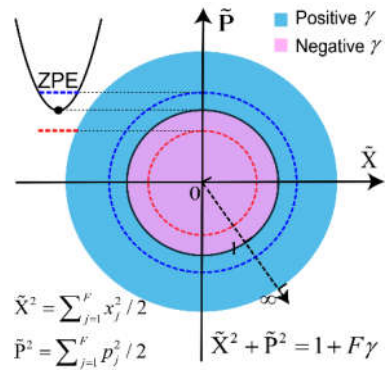

Nonadiabatic processes play an important role when two or more coupled electronic states are involved in complex systems in chemistry, biology, and materials science[1-3]. Of several useful theoretical frameworks for developing practical nonadiabatic dynamics methods, one is the mapping Hamiltonian model proposed in the pioneering work of Meyer and Miller[4]. It maps a coupled *F*-electronic-state Hamiltonian operator

$$\hat{H} = \sum_{n,m=1}^{F} H_{nm}(\mathbf{R},\mathbf{P})|n\rangle\langle m| \qquad (1)$$

onto a $2F$-dimensional Cartesian phase space $\{\mathbf{x},\mathbf{p}\}=\{x^{(1)},\ldots,x^{(F)},p^{(1)},\ldots,p^{(F)}\}$

$$H_{\mathrm{MM}}(\mathbf{x},\mathbf{p};\mathbf{R},\mathbf{P}) = \frac{1}{2}\mathbf{P}^{T}\mathbf{M}^{-1}\mathbf{P} + \sum_{n,m=1}^{F}\left[\frac{1}{2}(x^{(n)}x^{(m)}+p^{(n)}p^{(m)})-\gamma\delta_{nm}\right]V_{nm}(\mathbf{R}) \qquad (2)$$

in the diabatic representation, where $\{\mathbf{R},\mathbf{P}\}$ represents the coordinate and momentum variables for the nuclear degrees of freedom (DOFs) and $\gamma$ is the parameter accounting for the zero-point energy for each continuous electronic DOF. While $\gamma=1/2$ in Meyer and Miller's original version[4], the zero-point-energy (ZPE) parameter is set to $1/3$, $(\sqrt{3}-1)/2$, $(\sqrt{F+1}-1)/F$, and other non-negative values in its semiclassical/quasiclassical applications[5-16]. Since it is conventionally thought that the zero-point energy should be positive, to the best of our knowledge no negative value has ever been chosen for $\gamma$ since 1979.

In two alternative approaches, it has been shown that the Meyer-Miller mapping model can be derived in quantum mechanics[9, 17]. More interestingly, in the unified framework of phase space mapping models for the (coupled) multistate Hamiltonian (eq 1) in ref. [17], the mapping model reminiscent of the Meyer-Miller model reads

$$H_{\mathrm{MM}}(\mathbf{x},\mathbf{p};\mathbf{R},\mathbf{P}) = \sum_{n,m=1}^{F}\left[\frac{1}{2}(x^{(n)}x^{(m)}+p^{(n)}p^{(m)})-\gamma\delta_{nm}\right]H_{nm}(\mathbf{R},\mathbf{P}) \quad . \tag{3}$$

The comparison between eqs 3 and 2 (for kinetic energy term $\mathbf{P}^T\mathbf{M}^{-1}\mathbf{P}/2$) implies the constraint for the electronic DOFs[15]

$$\mathcal{S}(\mathbf{x},\mathbf{p}) : \sum_{n=1}^{F}\left[\frac{1}{2}\left((x^{(n)})^2+(p^{(n)})^2\right)\right]=1+F\gamma \quad . \tag{4}$$

Although we use the diabatic representation to reach eq 4, the constraint holds in the adiabatic representation as well. The constraint (eq 4) has already been hinted at in eqs 43 and 44 of ref. [17] and used in ref. [15] where $\gamma = 0$ is considered. The physical meaning of eq 4 requires only $\gamma > -1/F$. This confirms that negative values for the ZPE parameter, $\gamma$, are possible!

When the underlying mapping DOFs for each (electronic) state are considered to be those of a singly excited oscillator as first suggested in refs. [9] and [5], it is natural to view $\gamma$ as the ZPE parameter for the oscillator. In comparison, the derivation procedure of eq 3 in $F$-dimensional Hilbert space for the electronic DOFs in ref. [17] has interpreted that the physical concept of parameter $\gamma$ as a parameter for $\hat{\sigma}_z^{(n)}/2$, where $\hat{\sigma}_z$ is the Pauli spin matrix in the $z$ direction. It has also been clearly pointed out that parameter $\gamma$ can be negative[17].

We first establish the exact mapping with the constraint $\mathcal{S}(\mathbf{x},\mathbf{p})$ where $\gamma \in (-1/F,\infty)$ with a focus on the $F$ electronic DOFs. In $F$-dimensional Hilbert space with an orthogonal basis set $\{|n=1,\ldots,F\rangle\}$, the one-to-one correspondence between an operator and its phase space function reads

$$\hat{A} \mapsto A(\mathbf{x},\mathbf{p}) = \mathrm{Tr}_e\left[\hat{A}\hat{K}(\mathbf{x},\mathbf{p})\right] \tag{5}$$

and

$$A(\mathbf{x},\mathbf{p}) \mapsto \hat{A} = \int_{\mathcal{S}(\mathbf{x},\mathbf{p})} d\mu(\mathbf{x},\mathbf{p}) A(\mathbf{x},\mathbf{p}) \hat{K}^{-1}(\mathbf{x},\mathbf{p}) \ , \qquad (6)$$

where $\mathrm{Tr}_e[\cdots]$ represents the trace over the $F$ electronic states, $d\mu(\mathbf{x},\mathbf{p}) = F d\mathbf{x} d\mathbf{p}$ stands for an invariant measure over the constraint space $\mathcal{S}(\mathbf{x},\mathbf{p})$ of eq 4, the kernel is

$$\hat{K}(\mathbf{x},\mathbf{p}) = \sum_{n,m=1}^{F} \left[ \frac{1}{2} \left( x^{(n)} + ip^{(n)} \right) \left( x^{(m)} - ip^{(m)} \right) - \gamma \delta_{nm} \right] |n\rangle\langle m| \qquad (7)$$

and the inverse kernel is

$$\hat{K}^{-1}(\mathbf{x},\mathbf{p}) = \sum_{n,m=1}^{F} \left[ \frac{1+F}{2(1+F\gamma)^2} \left( x^{(n)} + ip^{(n)} \right) \left( x^{(m)} - ip^{(m)} \right) - \frac{1-\gamma}{1+F\gamma} \delta_{nm} \right] |n\rangle\langle m| \ . \qquad (8)$$

(The derivations of $\hat{K}(\mathbf{x},\mathbf{p})$ and $\hat{K}^{-1}(\mathbf{x},\mathbf{p})$ are presented in the Supporting Information.) As expected, the kernel and its inverse are properly normalized

$$\begin{array}{c} \mathrm{Tr}_e\left[\hat{K}(\mathbf{x},\mathbf{p})\right] = \mathrm{Tr}_e\left[\hat{K}^{-1}(\mathbf{x},\mathbf{p})\right] = 1 \\ \int_{\mathcal{S}(\mathbf{x},\mathbf{p})} d\mu(\mathbf{x},\mathbf{p}) \hat{K}(\mathbf{x},\mathbf{p}) = \int_{\mathcal{S}(\mathbf{x},\mathbf{p})} d\mu(\mathbf{x},\mathbf{p}) \hat{K}^{-1}(\mathbf{x},\mathbf{p}) = \hat{I}_e \end{array} \ , \qquad (9)$$

where $\hat{I}_e$ is the identity operator in $F$-dimensional Hilbert space for the electronic DOFs. We define the adjoint function in the electronic phase space for operator $\hat{B}$ as

$$\tilde{B}(\mathbf{x},\mathbf{p}) = \mathrm{Tr}_e\left[\hat{K}^{-1}(\mathbf{x},\mathbf{p})\hat{B}\right] \ . \qquad (10)$$

The one-to-one correspondence between the trace of a product of two operators and the overlap integral for the electronic DOFs is then

$$\mathrm{Tr}_e[\hat{A}\hat{B}] = \int_{\mathcal{S}(\mathbf{x},\mathbf{p})} d\mu(\mathbf{x},\mathbf{p}) A(\mathbf{x},\mathbf{p}) \tilde{B}(\mathbf{x},\mathbf{p}) \ . \qquad (11)$$

The time correlation function for two operators that involve *only* the electronic DOFs is

$$C_{AB}(t) = \mathrm{Tr}_e\left[\hat{A}(0)\hat{B}(t)\right] = \mathrm{Tr}_e\left[\hat{A}e^{i\hat{H}t/\hbar}\hat{B}e^{-i\hat{H}t/\hbar}\right] \qquad (12)$$

which becomes

$$C_{AB}(t) = \int_{\mathcal{S}(\mathbf{x},\mathbf{p})} d\mu(\mathbf{x},\mathbf{p}) A(\mathbf{x},\mathbf{p};0) \tilde{B}(\mathbf{x},\mathbf{p};t) \quad , \tag{13}$$

i.e., $\hat{A}$ or $\hat{B}$ is replaced by the corresponding Heisenberg operator in eq 11 as well as in eqs 5 and 10. Equation 13 can be evaluated as

$$\begin{aligned} C_{AB}(t) &= \int_{\mathcal{S}(\mathbf{x}_0,\mathbf{p}_0)} d\mu(\mathbf{x}_0,\mathbf{p}_0) A(\mathbf{x}_0,\mathbf{p}_0;0) \tilde{B}(\mathbf{x}(t),\mathbf{p}(t);0) \\ &= \int_{\mathcal{S}(\mathbf{x}_0,\mathbf{p}_0)} d\mu(\mathbf{x}_0,\mathbf{p}_0) A(\mathbf{x}_0,\mathbf{p}_0) \tilde{B}(\mathbf{x}_t,\mathbf{p}_t) \end{aligned} \tag{14}$$

along the trajectory $\{\mathbf{x}(t) \equiv \mathbf{x}_t, \mathbf{p}(t) \equiv \mathbf{p}_t\}$ with the initial phase point $\{\mathbf{x}_0,\mathbf{p}_0\}$ at time $0$. The equations of motion of the trajectory $\{\mathbf{x}_t,\mathbf{p}_t\}$ are given by the mapping Hamiltonian of eq 3, which is equivalent to solving the time-dependent Schrödinger equation for the electronic DOFs when the nuclear DOFs are frozen[4, 15, 17]. The mapping formulation (eqs 4-13) for the electronic DOFs is exact in quantum mechanics. Although the Meyer-Miller mapping Hamiltonian is used for demonstration, the mapping formulation can be applied to other mapping models (e.g., those of ref. [17]). The approach developed in ref. [15] is simply a specific case (where $\gamma = 0$) of the formulation.

In nonadiabatic systems, operators $\hat{A}$ and $\hat{B}$ often involve both electronic and nuclear DOFs. The trace operation in eq 12 should then be over both electronic and nuclear DOFs. Equation 13 can be extended as

$$C_{AB}(t) = \text{Tr}_{n,e}\left[\hat{A}(0)\hat{B}(t)\right] = \frac{1}{(2\pi\hbar)^N} \int d\mathbf{R}d\mathbf{P} \int_{\mathcal{S}(\mathbf{x},\mathbf{p})} d\mu(\mathbf{x},\mathbf{p}) A_W(\mathbf{R},\mathbf{P};\mathbf{x},\mathbf{p};0) \tilde{B}_W(\mathbf{R},\mathbf{P};\mathbf{x},\mathbf{p};t), \tag{15}$$

where the Wigner functions for the nuclear DOFs are

$$A_W(\mathbf{R},\mathbf{P};\mathbf{x},\mathbf{p}) = \int d\boldsymbol{\Delta} \left\langle \mathbf{R} - \frac{\boldsymbol{\Delta}}{2} \middle| A(\hat{\mathbf{R}},\hat{\mathbf{P}};\mathbf{x},\mathbf{p}) \middle| \mathbf{R} + \frac{\boldsymbol{\Delta}}{2} \right\rangle e^{i\boldsymbol{\Delta}\cdot\mathbf{P}/\hbar} \tag{16}$$

and

$$\tilde{B}_W(\mathbf{R},\mathbf{P};\mathbf{x},\mathbf{p}) = \int d\mathbf{\Delta} \left\langle \mathbf{R} - \frac{\mathbf{\Delta}}{2} \right| \tilde{B}(\hat{\mathbf{R}},\hat{\mathbf{P}};\mathbf{x},\mathbf{p}) \left| \mathbf{R} + \frac{\mathbf{\Delta}}{2} \right\rangle e^{i\mathbf{\Delta}\cdot\mathbf{P}/\hbar} \quad . \tag{17}$$

In eq 15, $\int d\mathbf{R} d\mathbf{P} \cdots$ is over the full Wigner phase space for the $N$ nuclear DOFs. On the right-hand side of eq 16 or eq 17 $\{\mathbf{x},\mathbf{p}\}$ is viewed as parameters.

$A(\hat{\mathbf{R}},\hat{\mathbf{P}};\mathbf{x},\mathbf{p})$ of eq 16 and $\tilde{B}(\hat{\mathbf{R}},\hat{\mathbf{P}};\mathbf{x},\mathbf{p})$ of eq 17 are produced by eq 5 for operator $\hat{A}$ and by eq 10 for operator $\hat{B}$, respectively, only for the electronic DOFs. Equation 15 offers an exact formulation as long as nuclear dynamics in eq 15 is also exactly solved for the coupled multielectronic-state Hamiltonian in eq 1, regardless of the choice of the ZPE parameter, $\gamma$, in domain $(-1/F, \infty)$ for the constraint space (eq 4).

It is often far from trivial to treat the nuclear DOFs in an exact fashion for general nonadiabatic systems. Employing the linearized semiclassical initial value representation (LSC-IVR)/classical Wigner approach to approximate nuclear dynamics[5], we obtain

$$C_{AB}(t) = \frac{1}{(2\pi\hbar)^N} \int d\mathbf{R}_0 d\mathbf{P}_0 \int_{\mathcal{S}(\mathbf{x}_0,\mathbf{p}_0)} d\mathbf{\mu}(\mathbf{x}_0,\mathbf{p}_0) A_W(\mathbf{R}_0,\mathbf{P}_0;\mathbf{x}_0,\mathbf{p}_0) \tilde{B}_W(\mathbf{R}_t,\mathbf{P}_t;\mathbf{x}_t,\mathbf{p}_t) \quad , \tag{18}$$

where trajectory $(\mathbf{R}_t,\mathbf{P}_t;\mathbf{x}_t,\mathbf{p}_t)$ follows classical Hamilton equations of motion yielded by the Meyer-Miller mapping Hamiltonian (eq 2) with initial total phase point $(\mathbf{R}_0,\mathbf{P}_0;\mathbf{x}_0,\mathbf{p}_0)$. The approach is denoted the extended classical mapping model (eCMM). For instance, consider that the initial total density is $\hat{\rho}_{nuc} \otimes |n\rangle\langle n|$ where $\hat{\rho}_{nuc}$ is the initial density operator for the nuclear DOFs and only state $|n\rangle$ is occupied at the beginning. The population of state $|m\rangle$ at time $t$ is

$$P_{m\leftarrow n}(t) = \text{Tr}\left[\hat{\rho}_{nuc} |n\rangle\langle n| e^{i\hat{H}t/\hbar} |m\rangle\langle m| e^{-i\hat{H}t/\hbar}\right] \quad . \tag{19}$$

Equation 18 leads to the phase space expression of eq 19

$$P_{m\leftarrow n}(t) = \frac{1}{(2\pi\hbar)^N} \int d\mathbf{R}_0 d\mathbf{P}_0 \int_{S(\mathbf{x}_0,\mathbf{p}_0)} d\mu(\mathbf{x}_0,\mathbf{p}_0) \left[\frac{1}{2}\left((x_0^{(n)})^2 + (p_0^{(n)})^2\right) - \gamma\right] \rho_W^{(nuc)}(\mathbf{R}_0,\mathbf{P}_0)$$
$$\times \left[\frac{1+F}{2(1+F\gamma)^2}\left((x_t^{(m)})^2 + (p_t^{(m)})^2\right) - \frac{1-\gamma}{1+F\gamma}\right] \quad , \quad (20)$$

where $\gamma \in (-1/F, \infty)$ and $\rho_W^{(nuc)}(\mathbf{R}_0,\mathbf{P}_0)$ is the Wigner function for $\hat{\rho}_{nuc}$. We emphasize that there exists only a single value for the ZPE parameter, $\gamma$, in the whole mapping scheme in the paper. That is, the ZPE parameter, $\gamma$, of the mapping Hamiltonian (eq 2 or 3) is the same as that of the constraint space (eq 4) as well as that employed in the expression of the time correlation function of nonadiabatic dynamics (eq 15 or 18).

It is worth mentioning that, independent of our work[15, 17], in the symmetrical quasiclassical Meyer-Miller (SQC/MM) approach[18] where the ZPE parameter in the Meyer-Miller Hamiltonian is modified on a per trajectory basis, a constraint of the classical actions has been considered such that the original Meyer-Miller Hamiltonian (eq 2) is consistent with the typically-used symmetrized form of the Hamiltonian. A generalized triangle windowing scheme is used for the initial and final conditions in the SQC/MM approach[18]. It will be interesting to determine whether a rigorous derivation for this trajectory-adjusted ZPE parameter approach is possible and whether the modification strategy on a per trajectory basis can be useful in our exact unified formulation.

Consider the spin-boson model that describes a two-state system coupled with a bosonic bath environment[19]. Such a prototype model in theoretical physics and chemistry includes key aspects of condensed phase nonadiabatic quantum systems[19-22]. Its Hamiltonian operator reads

$$\hat{H} = \varepsilon\hat{\sigma}_z + \Delta\hat{\sigma}_x + \left(\sum_j c_j \hat{R}_j\right)\hat{\sigma}_z + \sum_j \frac{1}{2}\left(\hat{P}_j^2 + \omega_j^2 \hat{R}_j^2\right) \quad , \quad (21)$$

where $\hat{\sigma}_x$ and $\hat{\sigma}_z$ are Pauli matrices in the $x$ and $z$ directions, respectively, $\varepsilon$ represents the detuning between states $|1\rangle$ and $|2\rangle$, $\Delta$ denotes the tunneling amplitude, and $\{\hat{R}_j, \hat{P}_j\}$ are the mass-weighted position and momentum operators of the $j$-th bath oscillator, respectively. Frequencies and coupling strengths $\{\omega_j, c_j\}$ are sampled from such as the Ohmic spectral density $J(\omega) = \frac{\pi}{2}\alpha\omega e^{-\omega/\omega_c}$, where $\alpha$ is the Kondo parameter and $\omega_c$ the cutoff frequency. The continuous spectral density is discretized into 300 effective bath modes to achieve full convergence. (See the Supporting Information for more details.)

The two-state system is assumed to be initially excited in state $|1\rangle$ with no correlation with the bosonic bath (i.e., the initial density is $|1\rangle\langle 1| \otimes e^{-\beta \hat{H}_b}/Z_b$, where $\hat{H}_b = \sum_j \frac{1}{2}\left(\hat{P}_j^2 + \omega_j^2 \hat{R}_j^2\right)$ is the bare bath Hamiltonian and $Z_b = \text{Tr}_n\left[e^{-\beta \hat{H}_b}\right]$ represents the partition function for the bath). While the initial distribution for the nuclear phase space is generated from the Wigner function of $e^{-\beta \hat{H}_b}/Z_b$, that for the electronic Cartesian phase space is uniformly sampled on constraint space $\mathcal{S}(\mathbf{x}_0, \mathbf{p}_0)$. The difference from the population of state $|1\rangle$ to that of state $|2\rangle$ (i.e., $D(t) = P_{1\leftarrow 1}(t) - P_{2\leftarrow 1}(t)$), can be evaluated from eq 20. We focus on more challenging asymmetric cases where Ehrenfest's mean field dynamics performs poorly. For fair comparison, the initial condition for the nuclear DOFs in the mean field dynamics is also sampled from the Wigner function of $e^{-\beta \hat{H}_b}/Z_b$, but the initial condition for the electronic DOFs for the two-state system is sampled from $\left((x^{(1)})^2 + (p^{(1)})^2\right)/2 = 1$ and $\left((x^{(2)})^2 + (p^{(2)})^2\right)/2 = 0$ as conventional Ehrenfest dynamics does. Five values for the ZPE parameter ($\gamma = -0.2$, $0$, $(\sqrt{3}-1)/2$, $0.5$, and $1$) are used for demonstration. Numerical simulations employ an ensemble of $10^5 - 10^6$ trajectories for convergence. Parameters for all examples of the spin-boson model are presented in atomic units (a. u.'s). For comparison, numerically exact results are obtained from the extended hierarchical equations of motion (eHEOM)[23-30], stochastic Liouville-von Neumann equation (SLNE)[31-32], and multilayer multiconfiguration time-dependent Hartree (ML-MCTDH)[33].

Figure 1 presents $D(t)$ for finite bath temperatures. The parameters range from high to low temperatures, from adiabatic to nonadiabatic domains, and from weak to strong system-bath coupling strengths. On the boundary of the adiabatic and nonadiabatic regions $(\omega_c = \Delta)$ and in the high-temperature region $(\beta\Delta = 0.25)$ the performance of eCMM is insensitive to the value for $\gamma$ (regardless of whether the ZPE parameter is negative or positive), of which the results perfectly match the exact data (Figures 1a,b). Figures 1c,d lies in the deeper nonadiabatic domain $(\omega_c = 2.5\Delta)$ with a relatively high temperature $(\beta\Delta = 0.25)$. All five values for $\gamma$ lead to nearly the same eCMM results that are almost identical to the exact data when the system-bath coupling $(\alpha = 0.1)$ is relatively weak (Figure 1c). As the system-bath coupling becomes stronger $(\alpha = 0.4)$, $\gamma = -0.2$ or 0 performs the best to describe the exact overdamped dynamics behavior, $\gamma = (\sqrt{3}-1)/2$ or 0.5 is slightly worse but produces results close to the exact ones, and $\gamma = 1$ exhibits a significant deviation (Figure 1d). Figure 1e,f demonstrates low-temperature $(\beta\Delta = 5)$ cases on the boundary of the adiabatic and nonadiabatic regions $(\omega_c = \Delta)$. In the weak dissipation case $(\alpha = 0.1)$ of Figure 1e, the underdamped oscillation is captured quite well by either $\gamma = -0.2$ or 0 in the whole range of $t\Delta = 15$. The results produced by $\gamma = (\sqrt{3}-1)/2$ or 0.5 deviate slightly from the exact data after $t\Delta = 5$, while $\gamma = 1$ considerably underestimates the oscillation amplitude. In the stronger dissipation case $(\alpha = 0.4)$ of Figure 1f, $\gamma = 1$ shows a more noticeable deviation as the time, $t\Delta$, increases, while the overall dynamic behavior is well reproduced by all four other values for $\gamma$. Figures 1g,h falls in the nonadiabatic domain $(\omega_c = 2.5\Delta)$ with a low

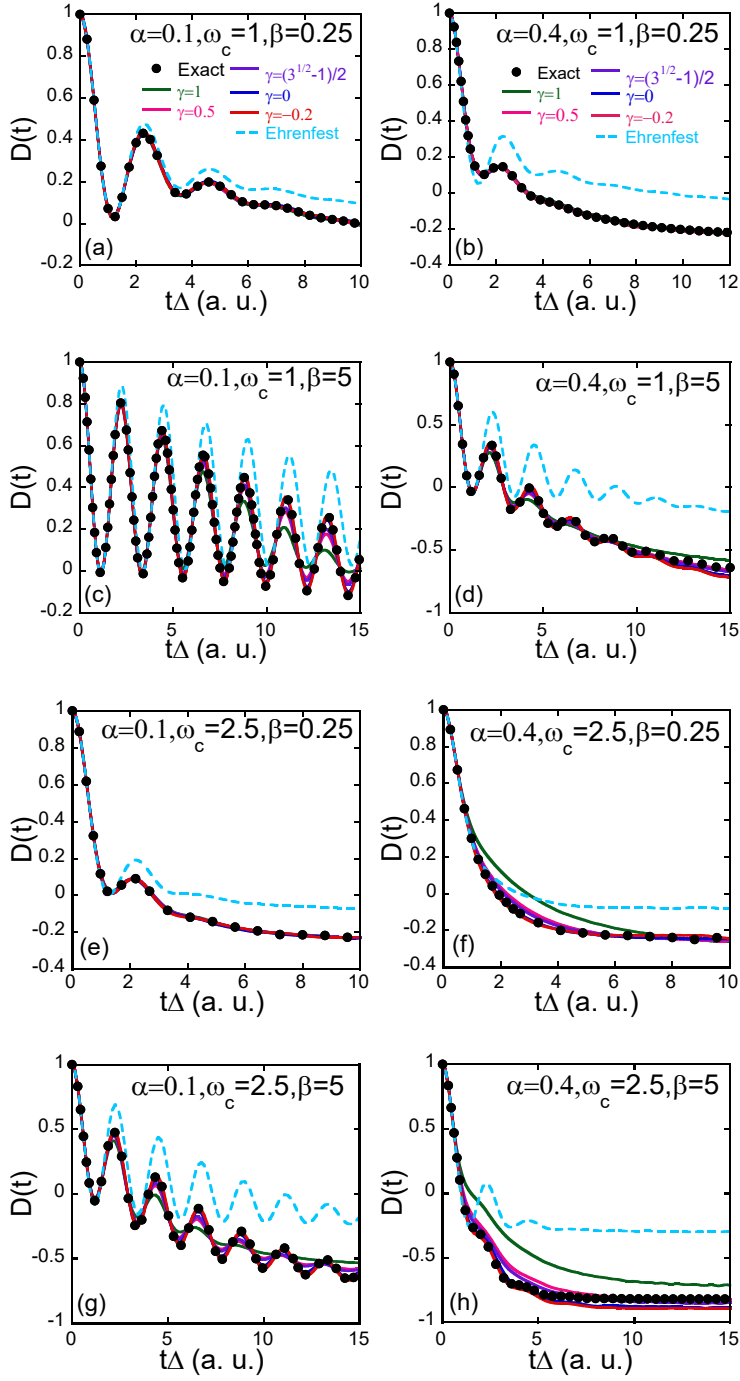

**Figure 1.** Population difference $D(t) = P_{1\leftarrow 1}(t) - P_{2\leftarrow 1}(t)$ of the spin-boson Hamiltonian with the Ohmic bath at finite temperature. The initial state is set to $|1\rangle$. In all eight panels $\varepsilon = \Delta = 1$. Black solid circles: numerically exact results in (a),(d), and (e)-(h) are from eHEOM; those in (c) and (d)

are from SLNE. Cyan dashed lines: Ehrenfest dynamics. Olive, magenta, purple, blue and red solid lines: eCMM approach with $\gamma=1$, $0.5$, $(\sqrt{3}-1)/2$, $0$, and $-0.2$, respectively.

temperature ($\beta\Delta = 5$). In the weak dissipation case in Figure 1g, while $\gamma = -0.2$ or $0$ is able to faithfully depict the distinct underdamped behavior in the exact time evolution of $D(t)$, the eCMM approach performs progressively worse in reproducing the long-time oscillation as the value of the ZPE parameter increases in the positive region. For example, $\gamma=1$ suffers a significant underestimation of the amplitude of oscillation after $t\Delta=5$. Figure 1h demonstrates the strong dissipation case instead. All values for $\gamma$ perform well for a short time before $t\Delta=1$. Three typical categories of results are observed for a longer time. $\gamma = -0.2$ and $0$ show the best performance during $t\Delta = 1-6$ but have a slight deviation from the asymptotic limit after $t\Delta=6$, $\gamma=(\sqrt{3}-1)/2$ and $0.5$ yield good estimations for the long-time equilibrium, and $\gamma=1$ predicts a slower decay rate and exhibits a large deviation after $t\Delta=1$, which performs substantially differently from the four other values.

Quantum dynamics of the dissipative two-state system at zero temperature is theoretically much more demanding[19]. Numerical methods for the spin-boson model at zero temperature often have severe slow convergence problems[21, 29, 34]. Figure 2 shows $D(t)$ for the four asymmetric spin-boson models with the Ohmic bath at zero temperature where exact benchmark results are available[29]. In the deeper nonadiabatic region $(\omega_c = 2.65\Delta)$ as shown in Figures 2a,b, $\gamma = -0.2$ and $0$ faithfully capture most dynamic behaviors, $\gamma=(\sqrt{3}-1)/2$ and $0.5$ lose the noticeable oscillation details in panel (a) and cause considerable deviation from the exact data in panel (b), while $\gamma=1$ demonstrates overall poor performance (except at short time) in both cases. In Figures 2c,d, the tunneling amplitude of the spin-boson model ($\Delta = 2$) is increased by a factor of 5 in comparison to that in Figures 2a,b. $\gamma = -0.2$ demonstrates the best performance in reproducing exact data, especially for the amplitude of dynamic oscillation in either case. The results of $\gamma = 0$ are very close to but slightly worse than those of $\gamma = -0.2$. As the positive value of the ZPE parameter becomes greater, the eCMM results increasingly deviate from exact data.

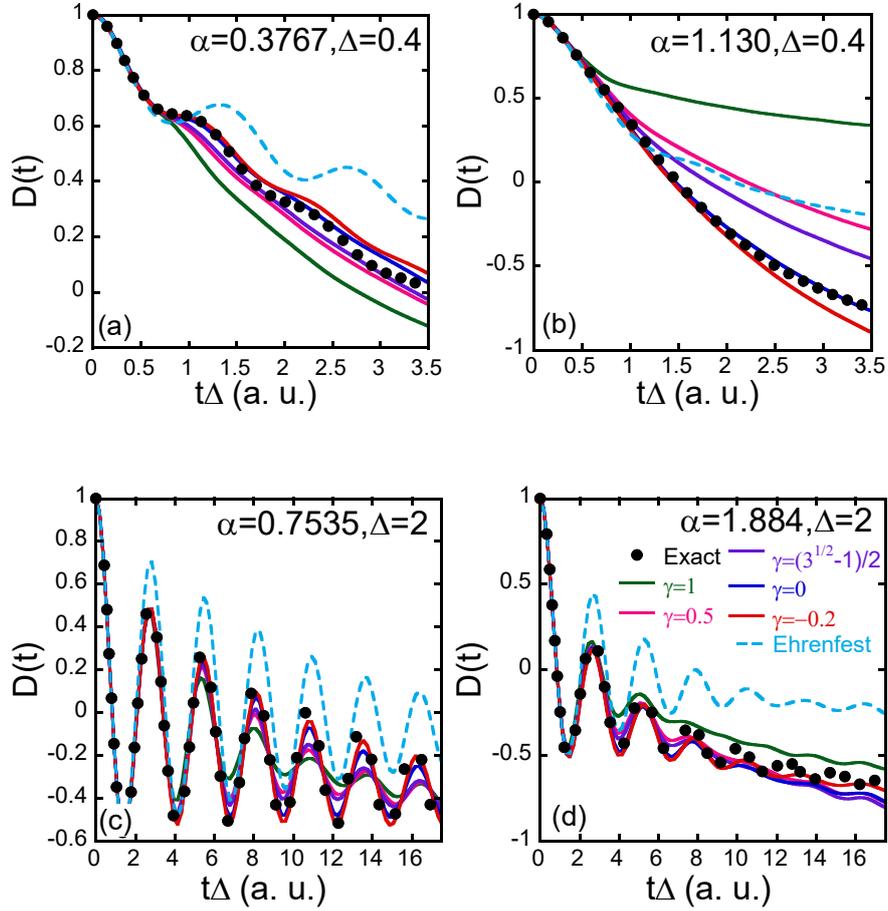

**Figure 2.** Population difference of the spin-boson Hamiltonian with the Ohmic bath at zero temperature. The initial state is set to $|1\rangle$. In all four panels $\omega_c = 1.0618$ and $\varepsilon = 1$. The other parameters are (a) $\{\alpha = 0.3767, \Delta = 0.4\}$, (b) $\{\alpha = 1.130, \Delta = 0.4\}$, (c) $\{\alpha = 0.7535, \Delta = 2\}$, (d)

$\{\alpha = 1.884, \Delta = 2\}$. The legend is the same as that in Figure 1, except exact results are produced by ML-MCTDH in ref. [29].

In all eCMM applications (with the Meyer-Miller mapping Hamiltonian) for the typical examples of the spin-boson model shown in the main text as well as in the Supporting Information, the negative ZPE parameter (e.g., $\gamma = -0.2$) practically faithfully depicts the dynamic oscillation behavior as well as the asymptotic long-time equilibrium, yielding reasonably accurate results in comparison to conventional positive ZPE parameters. In practice, the numerical performance is relatively insensitive to $\gamma$ in domain $\left(-1/F, 1/2\right]$ for *finite* temperature spin-boson systems. It is expected that $\gamma \in \left(-1/F, 1/2\right]$ may work reasonably well for other complex molecular systems.

In summary, a negative ZPE parameter is indeed possible and can be reasonably good for the Meyer-Miller mapping model for nonadiabatic molecular dynamics because the constraint implied in the kinetic energy term, $\mathbf{P}^T \mathbf{M}^{-1} \mathbf{P}/2$, in the mapping Hamiltonian requests only $\gamma \in \left(-1/F, \infty\right)$. When such a constraint is applied, we establish a novel and general formulation for constructing exact mapping models in the Cartesian phase space (for the electronic DOFs). Although the formulation is in principle exact irrespective of the meaningful value of $\gamma$, because the nuclear DOFs often cannot be treated in an exact fashion for general nonadiabatic systems, the performance of an approximated quantum dynamics method depends on the choice of the ZPE parameter. Although the dependence of $\gamma$ is relatively weak in many cases, it can become distinct in demanding regions (e.g., at low or zero temperature in condensed-phase systems). More insight will be warranted in the future to develop a novel strategy in the mapping formulation, which makes the numerical performance of practical nonadiabatic methods much less sensitive to the ZPE parameter, $\gamma$.

■ **ASSOCIATED CONTENT**

**Supporting Information**.

Supporting Information is available free of charge at

https://pubs.acs.org/doi/10.1021/acs.jpclett.1c00232.

> Kernel and inverse kernel in a unified mapping framework in quantum mechanics; Bath mode discretization; Numerical results for additional typical spin-boson models. (PDF)

■ **AUTHOR INFORMATION**

**Corresponding Author**

*E-mail: jianliupku@pku.edu.cn

**ORCID**

Xin He: 0000-0002-5189-7204

Zhihao Gong: 0000-0002-9643-633X

Baihua Wu: 0000-0002-1256-6859

Jian Liu: 0000-0002-2906-5858

**Author Contributions**

† H. X. and Z. G. contributed equally.

**Notes**

The authors declare no competing financial interest.

■ **Supporting Information**  https://pubs.acs.org/doi/10.1021/acs.jpclett.1c00232

**S1. The kernel and inverse kernel in a unified mapping framework in quantum mechanics**

**S1-A: Derivation of the kernel and inverse kernel**

Since Wigner's pioneering work in 1932[35], it has been well-known that a one-to-one correspondence mapping can be established from a quantum Hilbert space to a phase space, leading to phase space formulations of quantum mechanics[35-49]. Because it is natural in phase space formulations to obtain useful insight about the quantum-classical correspondence, they have been widely used in many areas of physics and physical chemistry.

Below we introduce a more general formulation for the one-to-one correspondence mapping in quantum mechanics. Such a formulation offers a natural mapping framework to include and go beyond the conventional unified classification scheme[15, 50] for phase space formulations.

The trace of a product of two operators is expressed in (mapping) phase space as

$$\mathrm{Tr}[\hat{A}\hat{B}] = \int d\mu(\mathbf{R},\mathbf{P}) A_C(\mathbf{R},\mathbf{P}) \tilde{B}_C(\mathbf{R},\mathbf{P}) \quad (S22)$$

where

$$A_C(\mathbf{R},\mathbf{P}) = \mathrm{Tr}\left[\hat{A}\hat{K}_C(\mathbf{R},\mathbf{P})\right]$$
$$\tilde{B}_C(\mathbf{R},\mathbf{P}) = \mathrm{Tr}\left[\hat{K}_C^{-1}(\mathbf{R},\mathbf{P})\hat{B}\right] \quad (S23)$$

and $d\mu(\mathbf{R},\mathbf{P})$ stands for the invariant measure over mapping phase space. The mapping kernel and its inverse satisfy the normalization

$$\mathrm{Tr}\left[\hat{K}_C(\mathbf{R},\mathbf{P})\right] = \mathrm{Tr}\left[\hat{K}_C^{-1}(\mathbf{R},\mathbf{P})\right] = 1 \quad (S24)$$

and

$$\int d\mu(\mathbf{R},\mathbf{P})\,\hat{K}_C(\mathbf{R},\mathbf{P}) = \int d\mu(\mathbf{R},\mathbf{P})\,\hat{K}_C^{-1}(\mathbf{R},\mathbf{P}) = \hat{I} \quad . \quad (S25)$$

Here, $\hat{I}$ is the identity operator in the corresponding space. The one-to-one correspondence mapping from $A_C(\mathbf{R},\mathbf{P})$ or $\tilde{B}_C(\mathbf{R},\mathbf{P})$ of Eq. S23 back to operator $\hat{A}$ or $\hat{B}$ is

$$\begin{aligned}\hat{A} &= \int d\boldsymbol{\mu}(\mathbf{R},\mathbf{P}) A_C(\mathbf{R},\mathbf{P}) \hat{K}_C^{-1}(\mathbf{R},\mathbf{P}) \\ \hat{B} &= \int d\boldsymbol{\mu}(\mathbf{R},\mathbf{P}) \tilde{B}_C(\mathbf{R},\mathbf{P}) \hat{K}_C(\mathbf{R},\mathbf{P})\end{aligned} \quad . \tag{S26}$$

When operators $\hat{A}$ and $\hat{B}$ are functions of the coordinate and momentum operators, the invariance measure in eqs S22, S25 and S26 can be chosen the simplest uniform version

$$d\boldsymbol{\mu}(\mathbf{R},\mathbf{P}) = (2\pi\hbar)^{-N} d\mathbf{R}d\mathbf{P} \quad , \tag{S27}$$

the kernel of mapping (in eq S23) is

$$\hat{K}_C(\mathbf{R},\mathbf{P}) = \left(\frac{\hbar}{2\pi}\right)^N \int d\boldsymbol{\zeta} \int d\boldsymbol{\eta}\, e^{i\boldsymbol{\zeta}\cdot(\hat{\mathbf{R}}-\mathbf{R})+i\boldsymbol{\eta}\cdot(\hat{\mathbf{P}}-\mathbf{P})} f(\boldsymbol{\zeta},\boldsymbol{\eta}) \quad , \tag{S28}$$

and its inverse is

$$\hat{K}_C^{-1}(\mathbf{R},\mathbf{P}) = \left(\frac{\hbar}{2\pi}\right)^N \int d\boldsymbol{\zeta} \int d\boldsymbol{\eta}\, e^{i\boldsymbol{\zeta}\cdot(\hat{\mathbf{R}}-\mathbf{R})+i\boldsymbol{\eta}\cdot(\hat{\mathbf{P}}-\mathbf{P})} \left[f(-\boldsymbol{\zeta},-\boldsymbol{\eta})\right]^{-1} \quad , \tag{S29}$$

where $f(\boldsymbol{\zeta},\boldsymbol{\eta})$ is a scalar function to determine the corresponding phase space, and $N$ is the total number of degrees of freedom (DOFs). Eqs S22-S29 are essential to reformulate the conventional unified classification scheme[15, 50]. Due to the Heisenberg uncertainty principle, such a one-to-one correspondence mapping is not unique in quantum mechanics. For example, the Wigner function[35-36] has

$$f(\boldsymbol{\zeta},\boldsymbol{\eta}) = 1 \quad , \tag{S30}$$

and the Husimi function[39] has

$$f(\zeta, \eta) = \exp\left(-\frac{\zeta^T \Gamma^{-1} \zeta}{4} - \frac{\hbar^2}{4}\eta^T \Gamma \eta\right) \quad . \tag{S31}$$

The integrals in eqs S22, S25, S26, S28 and S29 are over the whole phase space when the Wigner or Husimi function is employed.

When operators $\hat{A}$ and $\hat{B}$ are represented in $F$-dimensional Hilbert space, the essential equations of the mapping formulation are still eqs S22-S26, except the mapping kernel and its inverse are different. Consider operator $\hat{A} = \sum_{n,m=1}^{F} A_{nm} |n\rangle\langle m|$ in $F$-dimensional Hilbert space. Below we use $(\mathbf{x}, \mathbf{p})$ instead of $(\mathbf{R}, \mathbf{P})$ for mapping phase space for convenience. The one-to-one correspondence mapping between operator $\hat{A}$ and its Cartesian phase space function $A(\mathbf{x}, \mathbf{p})$ is chosen to be

$$A(\mathbf{x}, \mathbf{p}) = \text{Tr}\left[\sum_{m,n=1}^{F} A_{nm} |n\rangle\langle m| \hat{K}_C(\mathbf{x}, \mathbf{p})\right] = \sum_{m,n=1}^{F} A_{nm} \left[\frac{1}{2}(x_n + ip_n)(x_m - ip_m) - \gamma \delta_{nm}\right] , \tag{S32}$$

where $\{\mathbf{x}, \mathbf{p}\} = \{x_1, \ldots, x_F, p_1, \ldots, p_F\}$, the trace operation is over the $F$ Hilbert states, and $\gamma$ is a parameter. (Here we use $x_n$ instead of $x^{(n)}$ for convenience.) The mapping kernel $\hat{K}_C(\mathbf{x}, \mathbf{p})$ is presented in eq 7 of the paper. The normalization $\text{Tr}\left[\hat{K}_C(\mathbf{x}, \mathbf{p})\right] = 1$ of eq S24 requires

$$\text{Tr}\left[\hat{K}_C(\mathbf{x}, \mathbf{p})\right] = \sum_{k=1}^{F} \frac{1}{2}(x_k^2 + p_k^2) - F\gamma = 1 \quad , \tag{S33}$$

which defines the integral domain of the mapping phase space, i.e., $\mathcal{S}(\mathbf{x},\mathbf{p}): \sum_{k=1}^{F} \frac{1}{2}(x_k^2 + p_k^2) - F\gamma = 1$, such that eq S22 becomes eq 11 in the paper. The most straightforward way is to treat all phase variables equivalently such that the invariance measure is

$$d\boldsymbol{\mu}(\mathbf{x},\mathbf{p}) = \mu\, d\mathbf{x}d\mathbf{p} \;, \tag{S34}$$

with the scalar $\mu$ to be determined. The explicit expression of an integral in the defined domain, $\mathcal{S}(\mathbf{x},\mathbf{p})$, i.e., $\int_{\mathcal{S}(\mathbf{x},\mathbf{p})} \mu\, d\mathbf{x}d\mathbf{p}\, f(\mathbf{x},\mathbf{p})$, is

$$\begin{aligned}
\int_{\mathcal{S}(\mathbf{x},\mathbf{p})} \mu\, d\mathbf{x}d\mathbf{p}\, f(\mathbf{x},\mathbf{p}) &= \frac{\mu \int d\mathbf{x}d\mathbf{p}\, \delta\left(\frac{1}{2}\sum_{k=1}^{F}(x_k^2+p_k^2) - F\gamma - 1\right) f(\mathbf{x},\mathbf{p})}{\int d\mathbf{x}d\mathbf{p}\, \delta\left(\frac{1}{2}\sum_{k=1}^{F}(x_k^2+p_k^2) - F\gamma - 1\right)} \\
&= \frac{\mu\sqrt{2(1+F\gamma)}}{S_{2F-1}(\sqrt{2(1+F\gamma)})} \int d\mathbf{x}d\mathbf{p}\, \delta\left(\frac{1}{2}\sum_{k=1}^{F}(x_k^2+p_k^2) - F\gamma - 1\right) f(\mathbf{x},\mathbf{p})
\end{aligned} \tag{S35}$$

Here $S_{L-1}(R) = S_{L-1}(1) R^{L-1}$ is a measure of the ($L$-1)-dimensional constraint space defined in eq S44. (Some integration tricks are listed in Section S1-B.) Denote $\mathbf{X} = (\mathbf{x},\mathbf{p})$ for simplicity. It is not difficult to verify

$$\begin{aligned}
\int_{\mathcal{S}(\mathbf{x},\mathbf{p})} d\mathbf{X}\, (X_i X_j) &= \frac{1+F\gamma}{F}\delta_{ij} \\
\int_{\mathcal{S}(\mathbf{x},\mathbf{p})} d\mathbf{X}\, (X_i X_j X_k X_l) &= \frac{(1+F\gamma)^2}{F(F+1)}(\delta_{ij}\delta_{kl} + \delta_{ik}\delta_{jl} + \delta_{il}\delta_{jk})
\end{aligned} \tag{S36}$$

The first normalization of eq S25, i.e., $\int_{\mathcal{S}(\mathbf{x},\mathbf{p})} d\boldsymbol{\mu}(\mathbf{x},\mathbf{p})\, \hat{K}_C(\mathbf{x},\mathbf{p}) = \hat{I}$, leads to

$$\int_{\mathcal{S}(\mathbf{x},\mathbf{p})} \mu\, d\mathbf{x}d\mathbf{p} \sum_{n,m=1}^{F} \left[\frac{1}{2}(x_n + ip_n)(x_m - ip_m) - \gamma\delta_{nm}\right] |n\rangle\langle m| = \hat{I} \;. \tag{S37}$$

Substitution of eq S36 into eq S37 produces $\mu = F$, that is, $d\mu(\mathbf{x},\mathbf{p}) = F\, d\mathbf{x}d\mathbf{p}$ for eq S34.

Next we derive the inverse kernel, provided that the mapping kernel $\hat{K}_C(\mathbf{x},\mathbf{p})$ is a quadratic function of the phase variables as given by eq 7. Assume that the inverse mapping kernel $K_C^{-1}(\mathbf{x},\mathbf{p})$ takes the similar form

$$K_C^{-1}(\mathbf{x},\mathbf{p}) = \sum_{n,m=1}^{F} \left[ \frac{1}{2}\tilde{z}_1 (x_n + ip_n)(x_m - ip_m) - \tilde{z}_2 \delta_{nm} \right] |n\rangle\langle m| , \qquad (S38)$$

where $\tilde{z}_1$ and $\tilde{z}_2$ are two coefficients to be determined. Consider $\hat{A} = |m\rangle\langle n|$ and $\hat{B} = |k\rangle\langle l|$. The trace of the product is

$$\mathrm{Tr}[\hat{A}\hat{B}] = \delta_{nk}\delta_{ml} . \qquad (S39)$$

The mapping for the trace operation is

$$\mathrm{Tr}[\hat{A}\hat{B}] = \int_{S(\mathbf{x},\mathbf{p})} F\, d\mathbf{x}d\mathbf{p} \left[ \frac{1}{2}(x_n + ip_n)(x_m - ip_m) - \gamma \delta_{nm} \right] \\ \times \left[ \frac{1}{2}\tilde{z}_1 (x_l + ip_l)(x_k - ip_k) - \tilde{z}_2 \delta_{lk} \right]. \qquad (S40)$$

Integration in the right-hand side (RHS) of eq S40 and substitution of eq S39 into the left-hand side (LHS) of eq S40 yield

$$\delta_{nk}\delta_{ml} = \tilde{z}_1 \frac{(1+F\gamma)^2}{(F+1)} \delta_{nk}\delta_{ml} + \left[ \tilde{z}_1 \frac{(1+F\gamma)^2}{(F+1)} - \tilde{z}_2 - \tilde{z}_1\gamma - \tilde{z}_1 F\gamma^2 \right] \delta_{lk}\delta_{nm} . \qquad (S41)$$

This equation determines the two coefficients in the inverse mapping kernel

$$\tilde{z}_1 = \frac{1+F}{(1+F\gamma)^2},\quad \tilde{z}_2 = \frac{1-\gamma}{1+F\gamma} . \qquad (S42)$$

That is, substitution of eq S42 into eq S38 leads to the inverse kernel eq 8 of the paper.

Finally, we emphasize that it is straightforward to employ the procedure above to derive the kernel and its inverse for any other mapping models (e.g., those proposed in Ref. [17]) on constraint space for nonadiabatic systems. The unified mapping framework in quantum mechanics that we propose offers a powerful tool to consider both quantum systems represented in continuous coordinate space and those described in finite-dimensional Hilbert space. It is convenient to employ the unified mapping framework for systems where both continuous and discrete DOFs are involved.

**S1-B: Mathematical tricks for performing integrals on the constraint space**

Useful relations can be constructed between the $L$-dimensional normal distribution and uniform distribution on the constraint space. It is trivial to verify the equality

$$1 = (2\pi)^{-L/2} \int dX_1 dX_2 \cdots dX_L \exp\left[-\frac{1}{2}\sum_{k=1}^{L} X_k^2\right] = (2\pi)^{-L/2} \int_0^\infty d\xi e^{-\xi} \frac{S_{L-1}(\sqrt{2\xi})}{\sqrt{2\xi}} \quad , \tag{S43}$$

with the function $S_{L-1}(R)$ defined as

$$S_{L-1}(R) = R \int dX_1 dX_2 \cdots dX_L \delta\left(\frac{\sum_{k=1}^{L} X_k^2 - R^2}{2}\right) \quad . \tag{S44}$$

Scaling of $R$ in eq S44 yields the relation

$$S_{L-1}(R) = R^{L-1} S_{L-1}(1) \quad . \tag{S45}$$

Substitution of eq S45 into the RHS of eq S43 leads to

$$S_{L-1}(1) = \frac{2\pi^{L/2}}{\Gamma(L/2)} \quad , \tag{S46}$$

where $\Gamma(x) = \int_0^\infty e^{-t} t^{x-1} dt$ for $x > 0$ is the gamma function in mathematics and statistics.

Define $\mathbf{X} \equiv (X_1, X_2, \cdots, X_L)$ for convenience. The average in the ($L$-1)-dimensional constraint space $\mathcal{S}(\xi): \delta\left(\frac{1}{2}\sum_{k=1}^{L} X_k^2 - \xi\right)$ is denoted

$$\langle f \rangle_{\mathcal{S}(\xi)} = \frac{1}{S_{L-1}(\sqrt{2\xi})} \int d\mathbf{X}\, \delta\left(\frac{1}{2}\sum_{k=1}^{L} X_k^2 - \xi\right) f(\mathbf{X}) \;. \tag{S47}$$

The average of the $L$-dimensional independent identical normal distribution $X_k \sim \mathcal{N}(0,1)$ $(k=1,2,\cdots,L)$ is expressed as

$$\langle f \rangle_{\mathcal{N}_L(0,1)} = (2\pi)^{-L/2} \int d\mathbf{X}\, \exp\left[-\frac{1}{2}\sum_{k=1}^{L} X_k^2\right] f(\mathbf{X}) \;. \tag{S48}$$

Eqs S43, S47 and S48 generate the relation between the two averages

$$\langle f \rangle_{\mathcal{N}_L(0,1)} = (2\pi)^{-L/2} \int_0^\infty d\xi\, e^{-\xi}\, \frac{S_{L-1}(\sqrt{2\xi})}{\sqrt{2\xi}} \langle f \rangle_{\mathcal{S}(\xi)} \;. \tag{S49}$$

For example, considering the function $f(\mathbf{X}) = X_{n_1} X_{n_2} \cdots X_{n_k}$ that is a product of $k$ variables, after the scaling relation

$$\langle X_{n_1} X_{n_2} \cdots X_{n_k} \rangle_{\mathcal{S}(\xi)} = (2\xi)^{k/2} \langle X_{n_1} X_{n_2} \cdots X_{n_k} \rangle_{\mathcal{S}\left(\frac{1}{2}\right)} \tag{S50}$$

is employed, eq S49 yields

$$\langle X_{n_1} X_{n_2} \cdots X_{n_k} \rangle_{\mathcal{N}_L(0,1)} = (2\pi)^{-L/2} \int_0^\infty d\xi e^{-\xi} \frac{S_{L-1}(\sqrt{2\xi})}{\sqrt{2\xi}} \langle X_{n_1} X_{n_2} \cdots X_{n_k} \rangle_{\mathcal{S}(\xi)}$$

$$= \langle X_{n_1} X_{n_2} \cdots X_{n_k} \rangle_{\mathcal{S}\left(\frac{1}{2}\right)} S_{L-1}(1)(2\pi)^{-L/2} \int_0^\infty d\xi e^{-\xi} (2\xi)^{(L+k-2)/2} \quad . \quad (S51)$$

$$= \langle X_{n_1} X_{n_2} \cdots X_{n_k} \rangle_{\mathcal{S}\left(\frac{1}{2}\right)} \frac{2^{k/2} \Gamma(\frac{L+k}{2})}{\Gamma(L/2)}$$

The LHS of eq S51 can be formulated *via* Isserlis' theorem or Wick's probability theorem[51-52] to reach

$$\langle X_{n_1} X_{n_2} \cdots X_{n_k} \rangle_{\mathcal{N}_L(0,1)} = \sum \left( \prod \langle X_{n_i} X_{n_j} \rangle_{\mathcal{N}_L(0,1)} \right), \quad (S52)$$

where the notation $\sum(\ )$ represents the summation of all possible pair partitions for the $k$ variables $(X_{n_1}, X_{n_2}, \ldots, X_{n_k})$ and $\prod \cdots$ denotes the product of the $k/2$ pairs of $\langle X_{n_i} X_{n_j} \rangle$ ($k$ should be even for nonzero results). For instance,

$$\langle X_1 X_2 X_3 X_4 \rangle_{\mathcal{N}_L(0,1)} = \langle X_1 X_2 \rangle_{\mathcal{N}_L(0,1)} \langle X_3 X_4 \rangle_{\mathcal{N}_L(0,1)} + \langle X_1 X_3 \rangle_{\mathcal{N}_L(0,1)} \langle X_2 X_4 \rangle_{\mathcal{N}_L(0,1)}$$
$$+ \langle X_1 X_4 \rangle_{\mathcal{N}_L(0,1)} \langle X_2 X_3 \rangle_{\mathcal{N}_L(0,1)} \quad . \quad (S53)$$

Eqs S50, S51, and S52 suggest a convenient way for calculating $\langle X_{n_1} X_{n_2} \cdots X_{n_k} \rangle_{\mathcal{S}(\xi)}$ by using the averages of the *L*-dimensional independent identical normal distribution.

## S2. Bath Mode Discretization

The eCMM calculations for the spin-boson models involve harmonic modes obtained from the discretization procedure for the spectral density. The bath reorganization energy, $\lambda = \sum_i c_i^2 / 2\omega_i^2$, characterizes the averaged coupling strength between the bosonic bath and the spin system, of which the continuous form is $\lambda = 1/\pi \int d\omega J(\omega)/\omega$. The bath mode discretization is to equally divide $\lambda$ into $N_b$ segments. Each segment reads

$$\frac{c_j^2}{2\omega_j^2} = \frac{1}{\pi} \frac{J(\omega_j)}{\omega_j} \frac{1}{\rho(\omega_j)} , \qquad (S54)$$

where a number density function of frequency, $\rho(\omega)$, is introduced by $\int_0^{\omega_j} d\omega \rho(\omega) = j$ $(j = 1, \cdots, N_b)$. The discretization scheme thus depends on the formulation of $\rho(\omega)$. In our work, the number density function for Ohmic spectral density is set to be $\rho(\omega) \propto (N_b + 1) e^{-\omega/\omega_c}/\omega_c$, and each frequency and coupling strength satisfy[53-54]

$$\begin{cases} \omega_j = -\omega_c \ln\left[1 - j/(1+N_b)\right] \\ c_j = \omega_j \sqrt{\alpha \omega_c / (1+N_b)} \end{cases}. \qquad (S55)$$

Another typical kind of bath spectral density is the Debye spectral density, whose form is

$$J(\omega) = 2\lambda \frac{\omega_c \omega}{\omega_c^2 + \omega^2} . \qquad (S56)$$

In our simulations, eq S56 is discretized with $\rho(\omega) \propto (N_b + 1)/(\omega^2 + \omega_c^2)$, and each frequency and coupling strength follow[53]

$$\begin{cases} \omega_j = \omega_c \tan\left[\pi/2\left(1-j/(1+N_b)\right)\right] \\ c_j = \omega_j\sqrt{2\lambda/(1+N_b)} \end{cases}. \tag{S57}$$

According to our previous work[15], reliable results can be obtained with only $N_b = 50-100$ discretized harmonic modes. In the present work, we use $N_b = 300$ for all cases to guarantee full convergence.

### S3. Numerical Results for Additional Typical Spin-Boson Models

Numerical results for more typical examples of the spin-boson model are presented in Figures S1 and S2. Figure S1 includes eCMM results for the Ohmic bath in the nonadiabatic regime with relatively low temperatures. Figure S2 displays results for three typical spin-boson models with the Debye bath. These six additional typical examples have been investigated in Ref. [15].

**Figure S1** (Color). **Spin-boson models in the nonadiabatic region with relatively low temperature.** The parameters for the spin system are $\varepsilon = \Delta = 1$. The spectral density of the bosonic bath follows an Ohmic form. The bath parameters are (a) $\{\alpha = 0.4, \beta = 5, \omega_c = 2\}$, (b) $\{\alpha = 0.2, \beta = 5, \omega_c = 2.5\}$, and (c) $\{\alpha = 0.2, \beta = 10, \omega_c = 2.5\}$. In each figure, the black solid circles represent numerically exact results from eHEOM approach, while the red, blue, purple, magenta, olive solid lines are the predictions from eCMM under $\gamma = -0.2, 0, (\sqrt{3}-1)/2, 0.5$, and $1$.

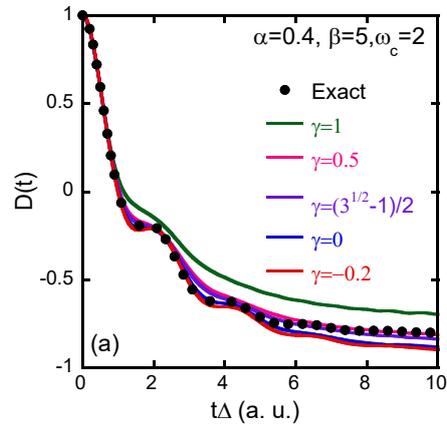

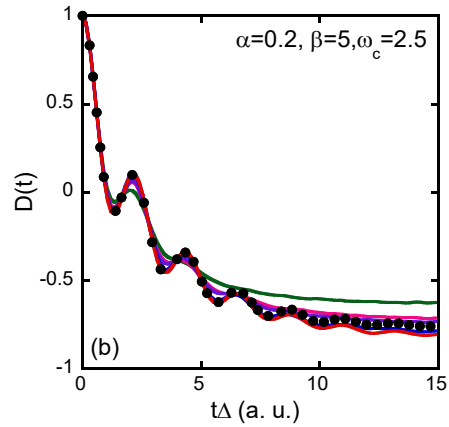

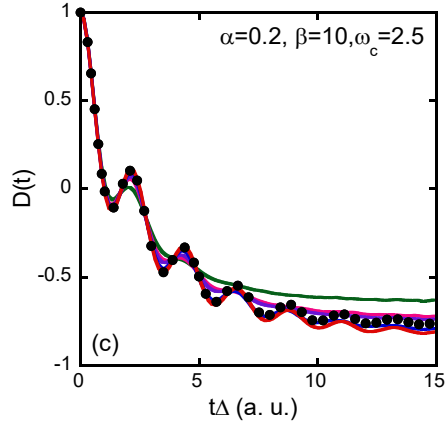

**Figure S2** (Color). **Spin-boson models with the Debye spectral density.** The parameters for the spin system are $\varepsilon = \Delta = 1$. The spectral density of the bosonic bath follows a Debye form. The bath parameters are (a) $\{\lambda = 0.25, \beta = 0.5, \omega_c = 0.25\}$, (b) $\{\lambda = 0.25, \beta = 0.5, \omega_c = 5\}$, and (c) $\{\lambda = 0.25, \beta = 50, \omega_c = 5\}$. Except that numerically exact results are calculated by the HEOM approach, the legend is the same as that in Figure S1.

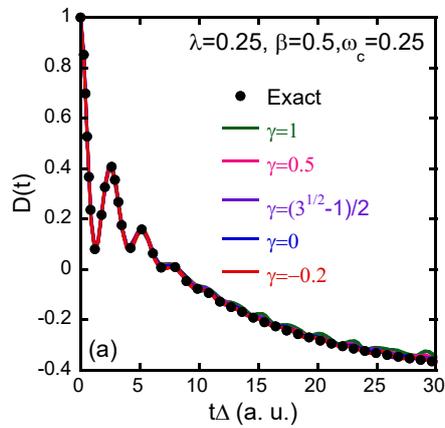

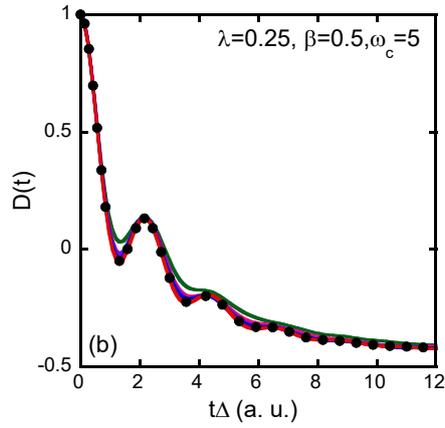

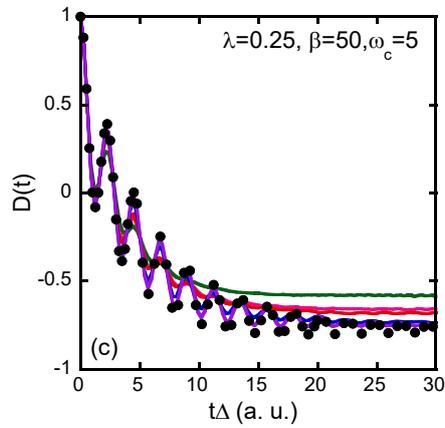

# ACKNOWLEDGMENT

This work was supported by the National Natural Science Foundation of China (NSFC) (grant no. 21961142017), and by the Ministry of Science and Technology of China (MOST) (grant no. 2017YFA0204901). We acknowledge the High-Performance Computing Platform of Peking University, Beijng PARATERA Tech CO.,Ltd., and the Guangzhou Supercomputer Center for providing computational resources. We thank Qianlong Wang and Jianlan Wu for providing the eHEOM results in Figure 1a,b,g,h (in the main text), Figure S1 and Figure S2b,c (in the Supporting Information).